\documentclass[aps,pre,english,showpacs,showkeys,preprintnumbers,onecolumn,floatfix,nofootinbib,10pt]{revtex4-2}
\usepackage{eurosym}
\usepackage{amssymb}
\usepackage{graphicx,color}
\usepackage{epsfig}
\usepackage{rotating}
\usepackage[T1]{fontenc}
\usepackage{latexsym,epsfig}
\usepackage[latin9]{inputenc}
\usepackage{amsfonts}
\usepackage{dcolumn}
\usepackage{bm}
\usepackage{babel}
\usepackage{hyperref}
\usepackage{amsmath,mathrsfs}

\begin{document}

\title{Exploring Transition from Stability to Chaos through Random Matrices}
\author{Roberto da Silva, Sandra Prado}
\affiliation{Instituto de F\'{\i}sica, Universidade Federal do Rio Grande do
Sul}

\begin{abstract}
This study explores the application of random matrices to track chaotic
dynamics within the Chirikov standard map. Our findings highlight the
potential of matrices exhibiting Wishart-like characteristics, combined with
statistical insights from their eigenvalue density, as a promising avenue
for chaos monitoring. Inspired by a technique originally designed for
detecting phase transitions in spin systems, we successfully adapt and apply
it to identify analogous transformative patterns in the context of the
Chirikov standard map. Leveraging the precision previously demonstrated in
localizing critical points within magnetic systems in our prior research,
our method accurately pinpoints the Chirikov resonance-overlap criterion for
the chaos boundary at $K\approx 2.43$, reinforcing its effectiveness.\newline
.
\end{abstract}

\maketitle

\section{Introduction}

Chaotic behavior plays a crucial role in contemporary physics \cite%
{Lichtenberg}, as the comprehension of non-determinism under initial
conditions arises in various contexts, including the stabilization of
seemingly simple mechanical systems, such as the inverted pendulum \cite%
{Peretti,Prado}.

The identification of chaos in specific Hamiltonian systems can be
accomplished using traditional methods; however, there is room for the
development of various alternatives. In contrast, the theory of random
matrices has provided a robust and potent toolkit for describing several
aspects of physical phenomena. This journey began with Wigner's pioneering
work, which explained the intricate distribution of energies in heavy nuclei 
\cite{Wigner,Wigner2}. Subsequently, Dyson, displaying remarkable foresight,
discerned that the joint distribution of eigenvalues in symmetric random
matrices, characterized by well-behaved matrix entries, behaves analogously
to a Coulomb gas of charged particles exhibiting logarithmic repulsion \cite%
{Dyson}.

In recent times, the authors of this study have recognized the potential of
a class of matrices known as Wishart-like matrices \cite{Wishart},
demonstrating their successful application in characterizing the critical
behavior of spin systems. This insight is revealed through the analysis of
the spectra of these matrices, as presented in our prior works \cite%
{RMT2023,RMT2023-2,RMT2023-3}.

The concept revolves around considering a specified number of time
evolutions of magnetization, acquired through a particular dynamics (e.g.,
Metropolis), as columns within matrices \cite{RMT2023}. These rectangular
matrices are then transformed into square matrices by multiplying the matrix
by its transpose, wherein the eigenvalues of these matrices offer insights
into the correlations among the time series data. Notably, phase transitions
are associated with deviations from the Marchenko-Pastur eigenvalue density,
which typically characterizes uncorrelated time series data \cite{Marcenko}

This paper aims to investigate the relationship between time series
generated through simple map iterations, exhibiting chaotic behavior, and
the spectral properties of Wishart-like matrices constructed from these
series. In essence, we seek to determine whether chaotic behavior is
discernible in these spectra, thereby offering an alternative avenue for the
study of chaotic phenomena

To accomplish this objective in our study, we have opted for the Chirikov
map iteration method \cite{Chirikov}. This method's origins trace back to a
particle subjected to the influence of a kicked potential, governed by a
time-dependent Hamiltonian:

\begin{equation}
\mathcal{H}(q,p,t)=\frac{p^{2}}{2m}+K\cos q\sum\limits_{n=0}^{\infty }\delta
(t-n)\ .
\end{equation}

he dynamics consist of a sequence of free propagations interspersed with
periodic kicks. The Hamiltonian equations yield:%
\begin{align}
\frac{dq}{dt}& =\frac{\partial \mathcal{H}}{\partial p}=\frac{p}{m} \\
&  \notag \\
\frac{dp}{dt}& =-\frac{\partial \mathcal{H}}{\partial p}=K\sin
q\sum\limits_{n=0}^{\infty }\delta (t-n)\   \notag
\end{align}

Hence, the Chirikov standard map, which preserves the area in the phase
space of the two canonical dynamical variables ($q$ and $p$), is defined as
follows:

\begin{align}
p_{n+1}& =p_{n}+K\sin q_{n} \\
q_{n+1}& =q_{n}+p_{n}\text{.}  \notag
\end{align}

We consider a unitary mass, $m=1$, for which the dynamics can be visualized
either within a cylinder by taking $q$ mod $2\pi $ or on a torus. In the
latter scenario, we take $q\ $mod $2\pi $ and $p$ mod $2\pi $.

From this interaction, we construct square matrices and monitor their
spectra as a function of $K$. Our findings indicate that the theoretical
conjectures for the chaotic boundaries appear to be reflected in the minimal
and maximal values of eigenvalue fluctuations (moments of eigenvalue
density).

In next section we present a brief tutorial about random matrices with
particular interest in Wishart-like matrices. We will show how the method
worked to find the critical behavior of Ising model and why it must works to
find the chaotic behavior of the Chirikov map. In section

In the upcoming section, we offer a concise tutorial on random matrices,
with a dedicated emphasis on Wishart-like matrices. We will delve into how
this method has effectively revealed the critical behavior of the Ising
model and why we hold the expectation that it will similarly shed light on
the chaotic behavior of the Chirikov map. Following that, in Section \ref%
{Sec:Results}, we present our primary findings. Lastly, we draw our study to
a close by summarizing our conclusions in Section \ref{Sec:Conclusions}.

\section{Wishart-Like Random Matrices: An Exploration of Their Novel
Application in Statistical Mechanics}

\label{Sec:Whishartlike}

The foundation of random matrices theory can be traced back to its inception
within the realm of nuclear physics, as E. Wigner \cite{Wigner,Wigner2}
pioneered its development to describe the intricate energy levels of heavy
nuclei. Wigner achieved this by representing the nucleus's Hamiltonian using
matrices with randomly distributed entries

When considering symmetric matrices ($h_{ij}=h_{ji}$) with well-behaved
entries, i.e., entries following a probability density function $f(h)$ such
that 
\begin{align}
\int_{-\infty }^{\infty }dh_{ij}f(h_{ij})h_{ij}& <\infty \text{,} \\
&  \notag \\
\int_{-\infty }^{\infty }dh_{ij}f(h_{ij})h_{ij}^{2}& <\infty  \notag
\end{align}%
of a matrix $H$, with dimensions $N\times N$, featuring independent entries,
and thus characterized by a joint distribution given as: 
\begin{equation}
\Pr (\prod_{i<j}h_{ij})=\prod_{i<j}f(h_{ij}).
\end{equation}

This leads to a joint eigenvalue distribution $P(\lambda _{1},...,\lambda
_{N})$, and its eigenvalue density is defined as follows:

\begin{equation}
\sigma (\lambda )=\int_{-\infty }^{\infty }...\int_{-\infty }^{\infty
}P(\lambda ,\lambda _{2},\lambda _{3},...,\lambda _{N})\text{,}
\end{equation}%
which, under the earlier-stated conditions for the matrix entries $h_{ji}$,
is universally characterized by the semi-circle law \cite%
{Mehta,Soshnikov1998}:\newline
\begin{equation}
\sigma (\lambda )=\left\{ 
\begin{array}{l}
\frac{1}{\pi }\sqrt{2N-\lambda ^{2}}\text{if\ }\lambda ^{2}<2N \\ 
\\ 
0\ \text{if }\lambda ^{2}\geq 2N%
\end{array}%
\right.  \label{Eq:wigner}
\end{equation}

In the particular context where $f(h_{ij})=\frac{e^{-h_{ij}^{2}/2}}{\sqrt{%
2\pi }}$, we can estabilish the Boltzmann weight as follows: 
\begin{equation*}
P(\lambda _{1},...,\lambda _{N})=C_{N}\exp \left[ -\frac{1}{2}%
\sum\limits_{i=1}^{N}\lambda _{i}^{2}+\sum\limits_{i<j}\ln \left\vert
\lambda _{i}-\lambda _{j}\right\vert \right] \text{,}
\end{equation*}%
where $C_{N}^{-1}$ $=\int_{0}^{\infty }...\int_{0}^{\infty }d\lambda
_{1}...d\lambda _{N}\exp [-\mathcal{H}(\lambda _{1}...\lambda _{N})]$
denotes the inverse of the normalization constant for a Coulomb gas with the
Hamiltonian:%
\begin{equation*}
\mathcal{H}(\lambda _{1}...\lambda _{N})=\frac{1}{2}\sum_{i=1}^{N}\lambda
_{i}^{2}-\sum_{i<j}\ln \left\vert \lambda _{i}-\lambda _{j}\right\vert
\end{equation*}%
operating at an inverse temperature $\beta ^{-1}=1$. The final term exhibits
logarithmic repulsion, akin to the conventional Wigner/Dyson ensembles, as
elucidated by Dyson \cite{Dyson}. Simultaneously, the first term exerts an
attractive influence. In the context of hermitian or symplectic entries, as
elucidated by Mehta \cite{Mehta}, the outcome remains comparable.
Specifically, it yields $P(\lambda _{1},...,\lambda _{N})=C_{N}^{\beta }\exp
(-\beta \mathcal{H)}$, with $\beta $ taking values of 2 and 4, resulting in
a consistently shared eigenvalue density \ref{Eq:wigner}.

Despite the apparent analogy, there is no immediate bridge between the
thermodynamics of a real-world system and the fluctuations observed in
random matrices generated from data originating from that very system.
However, when one delves deeper into the quest for correlations, this bridge
starts to materialize. Its comprehension holds the key to unlocking insights
into phase transitions and critical phenomena within Thermostatistics.

It's worth noting that nearly three decades before Wigner and Dyson's
groundbreaking work, Wishart \cite{Wishart} pioneered the analysis of
correlated time series. Rather than resorting to Gaussian or Unitary
ensembles, he delved into the realm of the Wishart ensemble. This ensemble
primarily deals with random correlation matrices, distinguishing it from the
conventional approaches of his contemporaries.

In recent contributions \cite{RMT2023}, we have explored this avenue by
introducing the magnetization matrix element $m_{ij}$ representing the
magnetization of the $j$th time series at the $i$th Monte Carlo (MC) step
within a system of $N=L^{d}$ spins. For simplicity in our investigations, we
adopted $d=2$, the minimum dimension where a phase transition occurs in the
simple Ising model. Additionally, we delved into the mean-field Ising model 
\cite{RMT2023-2}, maintaining the same total number of spins.

Here, $i=1,...,N_{MC}$, and $j=1,...,N_{sample}$. Consequently, the
magnetization matrix $M$ assumes dimensions $N_{MC}\times N_{sample}$. To
scrutinize spectral properties more effectively, we propose an intriguing
alternative: rather than analyzing $M$, we turn our attention to the square
matrix of dimensions $N_{sample}\times $ $N_{sample}$:

\begin{equation*}
G=\frac{1}{N_{MC}}M^{T}M\ ,
\end{equation*}%
resulting in $G_{ij}=\frac{1}{N_{MC}}\sum_{k=1}^{N_{MC}}m_{ki}m_{kj}$ a
matrix well-known as the Wishart matrix \cite{Wishart}. At this juncture,
rather than continuing with $m_{ij}$ it becomes more advantageous to operate
with the matrix $M^{\ast }$, whose elements are defined through the
customary variables:

\begin{equation*}
m_{ij}^{\ast }=\frac{m_{ij}-\left\langle m_{j}\right\rangle }{\sqrt{%
\left\langle m_{j}^{2}\right\rangle -\left\langle m_{j}\right\rangle ^{2}}},
\end{equation*}%
where: 
\begin{equation*}
\left\langle m_{j}^{k}\right\rangle =\frac{1}{N_{MC}}%
\sum_{i=1}^{N_{MC}}m_{ij}^{k}\ .
\end{equation*}

Thereby: 
\begin{equation}
\begin{array}{lll}
G_{ij}^{\ast } & = & \frac{1}{N_{MC}}\sum_{k=1}^{N_{MC}}\frac{%
m_{ki}-\left\langle m_{i}\right\rangle }{\sqrt{\left\langle
m_{i}^{2}\right\rangle -\left\langle m_{i}\right\rangle ^{2}}}\frac{%
m_{kj}-\left\langle m_{j}\right\rangle }{\sqrt{\left\langle
m_{j}^{2}\right\rangle -\left\langle m_{j}\right\rangle ^{2}}} \\ 
&  &  \\ 
& = & \frac{\left\langle m_{i}m_{j}\right\rangle -\left\langle
m_{i}\right\rangle \left\langle m_{j}\right\rangle }{\sigma _{i}\sigma _{j}}%
\end{array}
\label{Eq:Correlation}
\end{equation}%
where $\left\langle m_{i}m_{j}\right\rangle =\frac{1}{N_{MC}}%
\sum_{k=1}^{N_{MC}}m_{ki}m_{kj}$ and $\sigma _{i}=\sqrt{\left\langle
m_{i}^{2}\right\rangle -\left\langle m_{i}\right\rangle ^{2}}$.
Analytically, when $m_{ij}^{\ast }$ are uncorrelated random variables, the
joint distribution of eigenvalues can be described by the Boltzmann weight 
\cite{GuhrPR,Seligman3}: 
\begin{equation*}
P(\lambda _{1},...,\lambda _{N_{sample}})=C_{N_{sample}}\exp [-\frac{%
N_{MC}^{2}}{2N_{sample}}\sum_{i=1}^{N_{sample}}\lambda _{i}-\frac{1}{2}%
\sum_{i=1}^{N_{sample}}\ln \lambda _{i}+\sum_{i<j}\ln \left\vert \lambda
_{i}-\lambda _{j}\right\vert ]\bigskip
\end{equation*}%
where $C_{N_{sample}}^{-1}=\int_{0}^{\infty }...\int_{0}^{\infty }d\lambda
_{1}...d\lambda _{N_{sample}}\exp [-\mathcal{H}(\lambda _{1}...\lambda
_{N_{sample}})]$, and this corresponds to the Hamiltonian:%
\begin{equation*}
\mathcal{H}(\lambda _{1}...\lambda _{N_{sample}})=\frac{N_{MC}^{2}}{%
2N_{sample}}\sum_{i=1}^{N_{sample}}\lambda _{i}+\frac{1}{2}%
\sum_{i=1}^{N_{sample}}\ln \lambda _{i}-\sum_{i<j}\ln \left\vert \lambda
_{i}-\lambda _{j}\right\vert \text{.}
\end{equation*}%
In this case, the density of eigenvalues $\rho (\lambda )$ of the matrix $%
G^{\ast }=\frac{1}{N_{MC}}M^{\ast T}M^{\ast }$ follows the well-known
Marcenko-Pastur distribution \cite{Marcenko}, which for our case we write as:

\begin{equation}
\rho (\lambda )=\left\{ 
\begin{array}{l}
\dfrac{N_{MC}}{2\pi N_{sample}}\dfrac{\sqrt{(\lambda -\lambda _{-})(\lambda
_{+}-\lambda )}}{\lambda }\ \text{if\ }\lambda _{-}\leq \lambda \leq \lambda
_{+} \\ 
\\ 
0\ \ \ \ \text{otherwise,}%
\end{array}%
\right.   \label{Eq:MP}
\end{equation}%
where $\lambda _{\pm }=1+\frac{N_{sample}}{N_{MC}}\pm 2\sqrt{\frac{N_{sample}%
}{N_{MC}}}.$

In our studies \cite{RMT2023,RMT2023-2}, we examined the behavior of $\rho _{%
\text{numerical}}(\lambda )$ by analyzing $m_{ij}$ data obtained from both a
two-dimensional Ising model and a mean-field Ising model. These models were
simulated at various temperatures using the single-spin flip Metropolis
dynamics.

In the first scenario, we considered square lattices with a linear dimension 
$L=100$, resulting in a total of $N=10000$ spins. We maintained the same
number of spins in the second case. Our simulations employed $N_{MC}=300$
and $N_{sample}=100$, which is computationally highly efficient.

We repeated the process $N_{run}=1000$ times to generate a sufficient number
of eigenvalues for constructing histograms and calculating numerical moments:

\begin{equation}
\left\langle \lambda^{k}\right\rangle _{\text{numerical}}=\frac{\sum
\limits_{i=1}^{N_{bins}}\lambda_{i}^{k}\rho_{\text{numerical}}(\lambda_{i})}{%
\sum\limits_{i=1}^{N_{bins}}\rho_{\text{numerical}}(\lambda_{i})}\text{,}
\label{Eq:fluctuations}
\end{equation}

Our histograms were constructed with $N_{bins}=100$. We antecipate that $%
\rho _{\text{numerical}}(\lambda )$ should approach $\rho (\lambda )$
according to Eq. \ref{Eq:MP} as $T\rightarrow \infty $ (in the paramagnetic
phase). In this situation $\left\langle \lambda ^{k}\right\rangle _{\text{%
numerical}}$ should closely align with theoretical value:%
\begin{equation}
\left\langle \lambda ^{k}\right\rangle =\int\limits_{-\infty }^{\infty
}\lambda ^{k}\rho (\lambda )d\lambda =\sum\limits_{j=0}^{k-1}\frac{\left( 
\frac{N_{sample}}{N_{MC}}\right) ^{j}}{j+1}\binom{k}{j}\binom{k-1}{j}\text{.}
\label{Eq:Exact_fluctuations}
\end{equation}%
For $k=1$, $\left\langle \lambda \right\rangle =1$, and we expect $%
\left\langle \lambda \right\rangle _{\text{numerical}}$ $\approx 1$ as $%
T\rightarrow \infty $. However for $T\approx T_{C}$ or $T<T_{C}$ the results
warrant closer examination. We present the results for fluctuations in the
Ising model, both in the two-dimensional and mean-field approximations, as
functions of temperature, employing the parameters described above, as
detailed in our previous observations \cite{RMT2023,RMT2023-2}. Figure \ref%
{Fig:Ising} vividly illustrates that fluctuations exhibit a distinctive
response concerning critical phenomena in the Ising model. This phenomenon
holds true irrespective of whether we're considering Monte Carlo simulations
in the two-dimensional Ising model or its mean-field approximation.

\begin{figure}[tbp]
\begin{center}
\includegraphics[width=0.6\columnwidth]{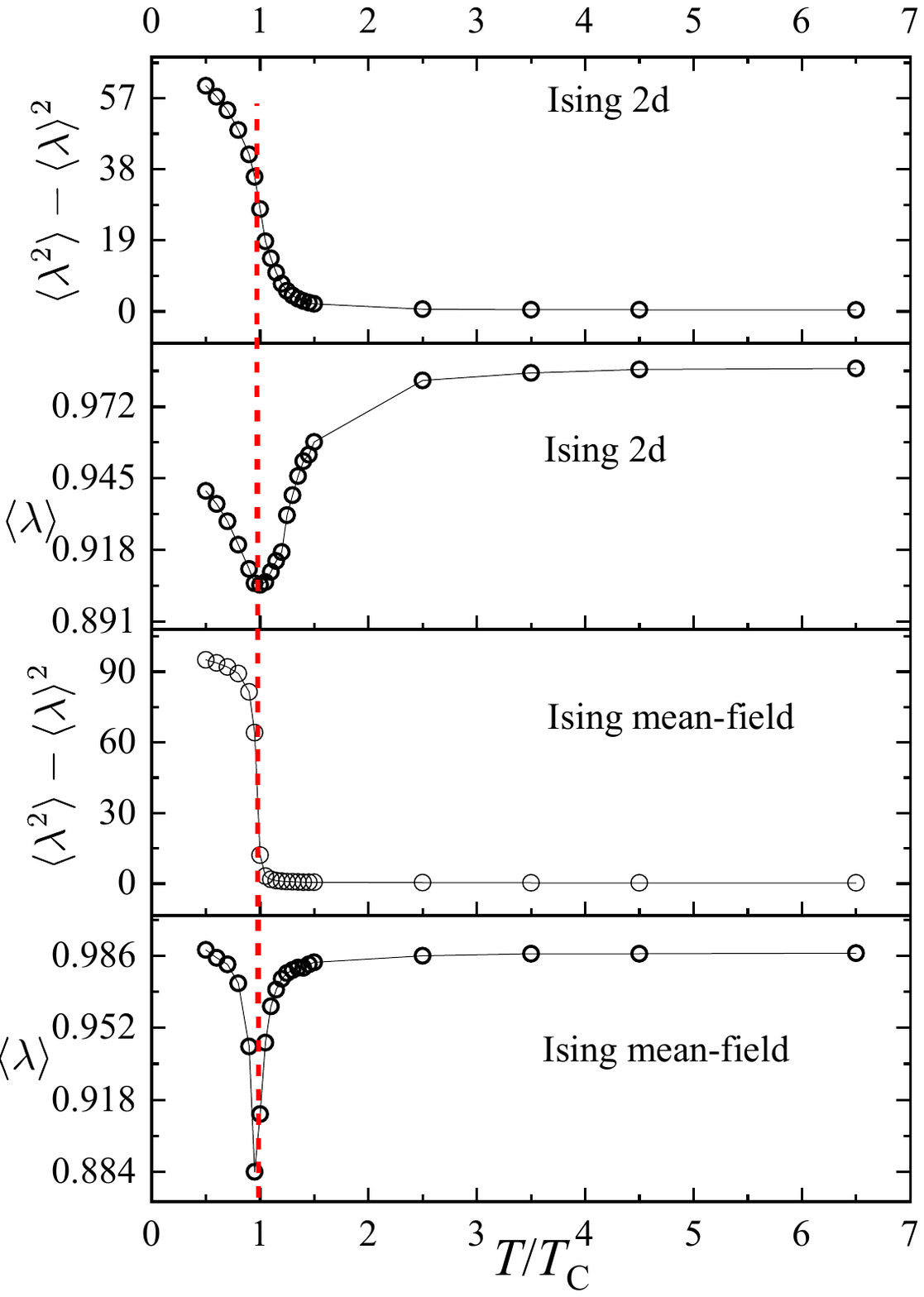}
\end{center}
\caption{The fluctuations of eigenvalues in $G$ were examined, taking into
account the Ising model in two settings: a two-dimensional square lattice
and a mean-field approximation, as reported in \protect\cite%
{RMT2023,RMT2023-2}. Notably, we can discern a distinct peak occurring at $%
T=T_{C}$ for the average eigenvalue in both formulations of the Ising model.
Additionally, an inflection point becomes apparent in the variance of the
eigenvalue.}
\label{Fig:Ising}
\end{figure}

We can clearly discern a minimum point in the behavior of $\left\langle
\lambda \right\rangle $ and an inflection point in $\left\langle \lambda
^{2}\right\rangle -\left\langle \lambda \right\rangle ^{2}$ occurring
precisely at $T=T_{C}$. As discussed in detail in \cite{RMT2023,RMT2023-2},
this phenomenon is closely tied to the emergence of a gap when $T<T_{C}$,
which subsequently closes as $T$ approaches $T_{C}$. For a more
comprehensive understanding, please refer to \cite{RMT2023,RMT2023-2}, and
also \cite{RMT2023-3}. It is worth noting that $\left\langle \lambda
\right\rangle _{\text{numerical}}$ approximates 1 for large values of $T$.

However, an intriguing question arises: How do $\left\langle \lambda
\right\rangle $ and $\left\langle \lambda ^{2}\right\rangle -\left\langle
\lambda \right\rangle ^{2}$ behave when considering iterations of the
Chirikov map instead of time-series data from spin systems? In the following
section, we present the key findings of this study.

\section{Main Results}

\label{Sec:Results}

We conducted iterations of the Chirikov map, mirroring the approach employed
in our study of the Ising model. In this case, we obtained matrix elements $%
m_{ij}$, which can now represent $q_{ij}$ (iterations for position
coordinates) or $p_{ij}$ (iterations for moments). Here, $i=1,...,N=200$
iteration steps, and $j$ ranges from 1 to $N_{sample}=100$ different series.
To initiate this process, we initialized random values for $q_{0}$ within
the range $[0,2\pi ]$ and $p_{0}$ within the range $[0,2\pi ]$.

To provide a pedagogical visualization of the map iteration, let's consider
a simple example. We observe the time evolutions $q_{i}$ and $p_{i}$ as
functions of step $i=1,2,...100$ for three different initial conditions: $j=1
$ ($p_{0}=0.1$ and $q_{0}=\pi $), $j=2$ ($p_{0}=1$and $q_{0}=1$), and $j=3$ (%
$p_{0}=\pi \ $and $q_{0}=0.1$). These evolutions are presented in Figure \ref%
{Fig:evolutions}, with the sequences displayed from bottom to top.

\begin{figure}[tbp]
\begin{center}
\includegraphics[width=1.0\columnwidth]{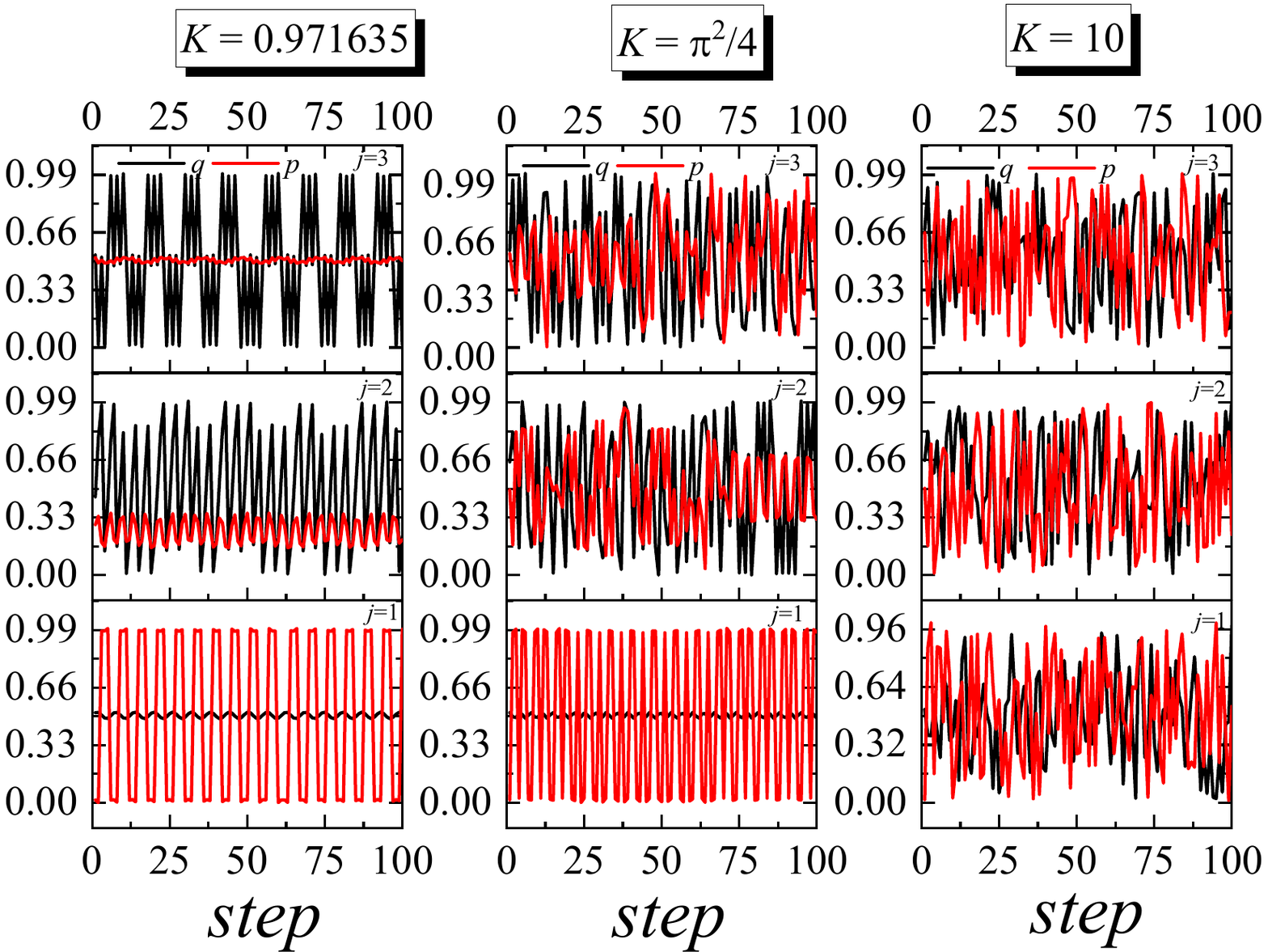}
\end{center}
\caption{We engaged in iterations of moment and position coordinates within
the framework of Chirikov's map. The columns of matrix $M$ were populated
with $N_{sample}$ iterations of position and moment pairs $q_{ij}$, $p_{ij}$%
. To illustrate this, we present a sample of time evolutions originating
from three distinct initial conditions $(p_{0},q_{0})=(0.1,\protect\pi )$,$%
(1,\protect\pi )$, and ($\protect\pi ,0.1$), respectively, displayed from
bottom to top. These evolutions were examined under three different
parameter values: $K=0.971635$ (signifying the destruction of the golden KAM
curve), $\frac{\protect\pi ^{2}}{4}$ (corresponding to the Chirikov criteria
for the chaos border), and $10$ (indicative of a state of complete chaos).}
\label{Fig:evolutions}
\end{figure}

For the sake of illustration, we selected three distinct values of $K$. The
first, $K=0.971635$, signifies a scenario where the golden KAM curve is
theoretically destroyed. The second, $K=\frac{\pi ^{2}}{4}$, aligns with the
Chirikov resonance-overlap criterion for defining the chaos border. Lastly,
we considered $K=10$, representing a situation characterized by complete and
unbridled chaos.

To enhance the illustrative aspect, we also generated Poincar� sections
corresponding to these three parameters, primarily for pedagogical purposes.
To create these sections, we explored values for $(N_{p}+1)(N_{q}+1)$
different initial conditions parametrized as follows: $q_{0}=\frac{2\pi }{%
N_{q}}l$ and $p_{0}=\frac{2\pi }{N_{p}}l$, where $l=0,1...,N_{q}$ and $%
m=0,1...,N_{p}$. For effective visualization, we employed $N_{q}=N_{p}=n=20$%
, resulting in $q_{0}$ and $p_{0}$ residing within the range $[0,2\pi ]$.

\begin{figure}[tbp]
\begin{center}
\includegraphics[width=1.0\columnwidth]{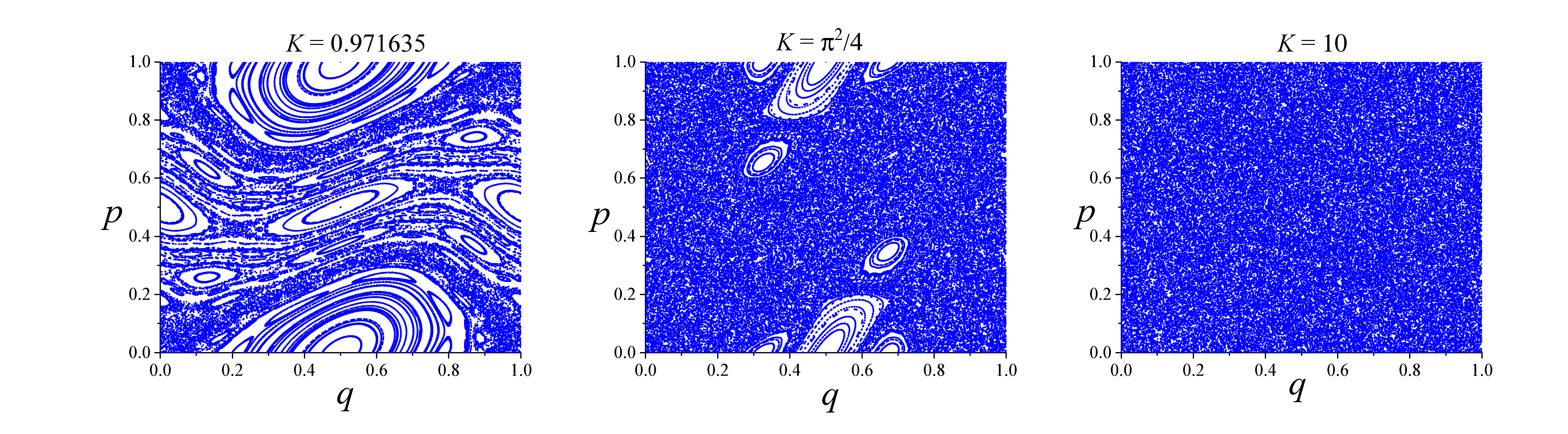}
\end{center}
\caption{Poincar� sections were constructed for the same set of $K$ values
employed in Figure \protect\cite{Fig:evolutions}.}
\label{Fig:Poincare}
\end{figure}

The intriguing and anticipated patterns can be vividly appreciated in Figure %
\ref{Fig:Poincare}. Following this pedagogical exploration of the Chirikov
method, we now proceed to utilize the random matrix method proposed here to
shed light on the chaos border.

Our algorithm constructs an ensemble of matrices, comprising $N_{run}=1000$
distinct matrices $G$ of dimensions $N_{sample}\times N_{sample}$. These
matrices correspond to $N_{run}$ varying initial conditions, which are
randomly selected with $q_{0},p_{0}\in \lbrack 0,2\pi ]$. Subsequently, we
diagonalize these matrices and organize the eigenvalues within the interval $%
\lambda _{\min }^{(\text{Numerical})}$ to $\lambda _{\max }^{(\text{Numerical%
})}$. We maintain a fixed number of bins, $N_{bin}=100$, and generate
histograms to calculate $\rho _{\text{numerical}}(\lambda _{i})$.

We carried out this process for various values of $K$, ranging from $K_{\min
}=0$ to $K_{\max }=10$. Consequently, we present the numerical density of
eigenvalues for three distinct $K$ values in Figure \ref%
{Fig:Density_of_eigenvalues}. In this figure, we display both the
eigenvalues of $G$ derived from the time evolutions of $q$ and $p$. It's
noteworthy that for small values of $K$, a noticeable difference is observed
when compared to the MP-law density, as described by equation \ref{Eq:MP}.
However, for $K=10$, a substantial match between the numerical results and
the theoretical prediction (MP-law, see Eq. \ref{Eq:MP}) is evident. It's
important to note that this match is not perfect, which aligns with
expectations since a perfect match would typically occur only for entirely
random time series, not those displaying complete chaos.

\begin{figure}[tbp]
\begin{center}
\includegraphics[width=0.8\columnwidth]{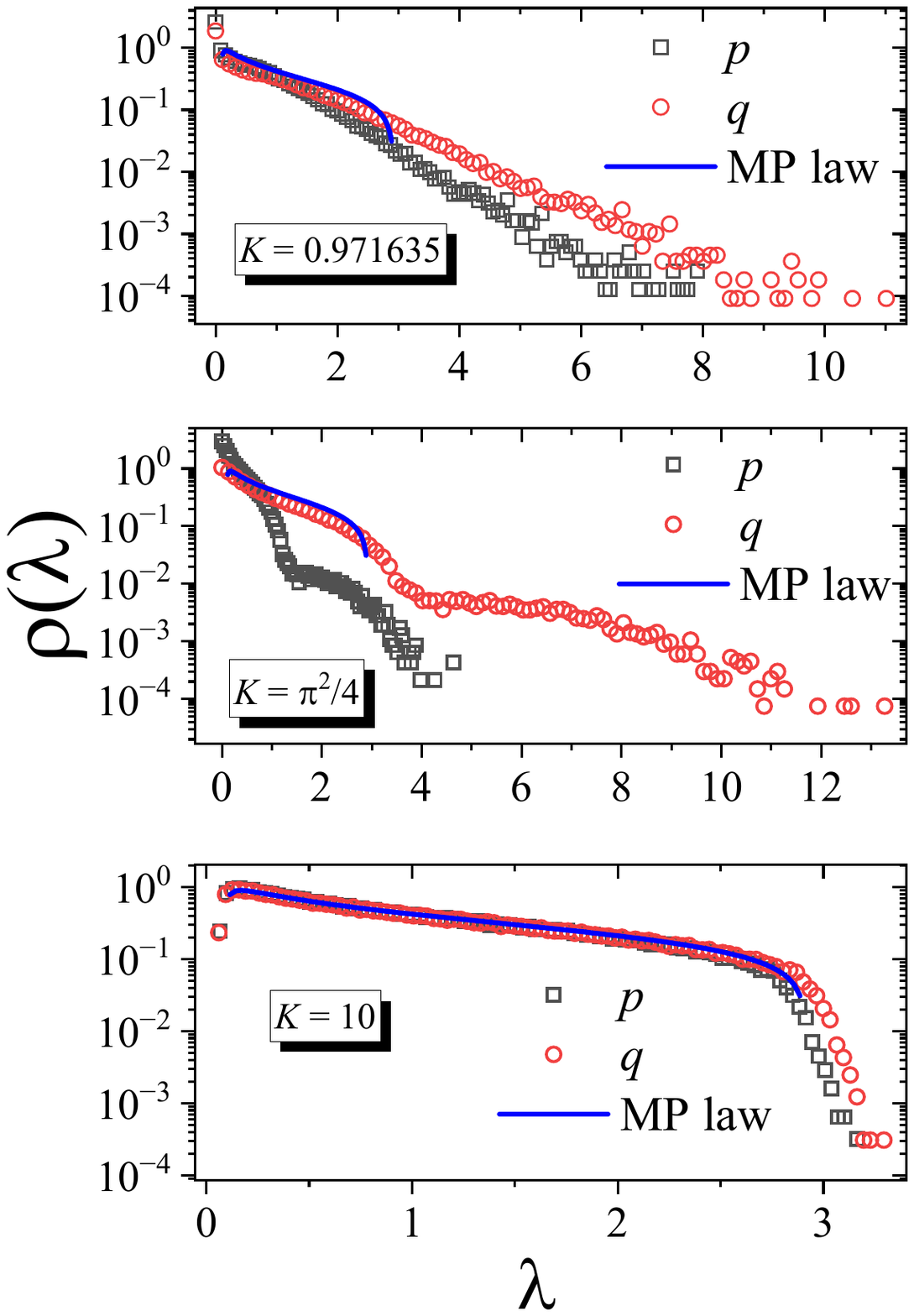}
\end{center}
\caption{The density of eigenvalues for various $K$ values is depicted here.
In the chaotic regime, we can discern a notable alignment with the MP-law,
indicating a good fit.}
\label{Fig:Density_of_eigenvalues}
\end{figure}

This suggests a potential approach for distinguishing chaos from random
behavior, a crucial point as highlighted in \cite{Massoler}. Our exploration
of random matrices holds promise in addressing this challenge, as
underscored by parallel research efforts in \cite{RMTFUTURA}.

In the context of spin systems (as discussed in \cite%
{RMT2023,RMT2023-2,RMT2023-3}), we have observed a restoration of the
Marchenko-Pastur law at elevated temperatures. While the system is
predominantly stochastic rather than highly chaotic in this scenario, we can
draw a meaningful analogy. To further investigate this phenomenon, we
examine the fluctuations of eigenvalues, as described in Equation\ref%
{Eq:fluctuations}, particularly those derived from time-evolutions of
moments.

\begin{figure}[tbp]
\begin{center}
\includegraphics[width=1.0\columnwidth]{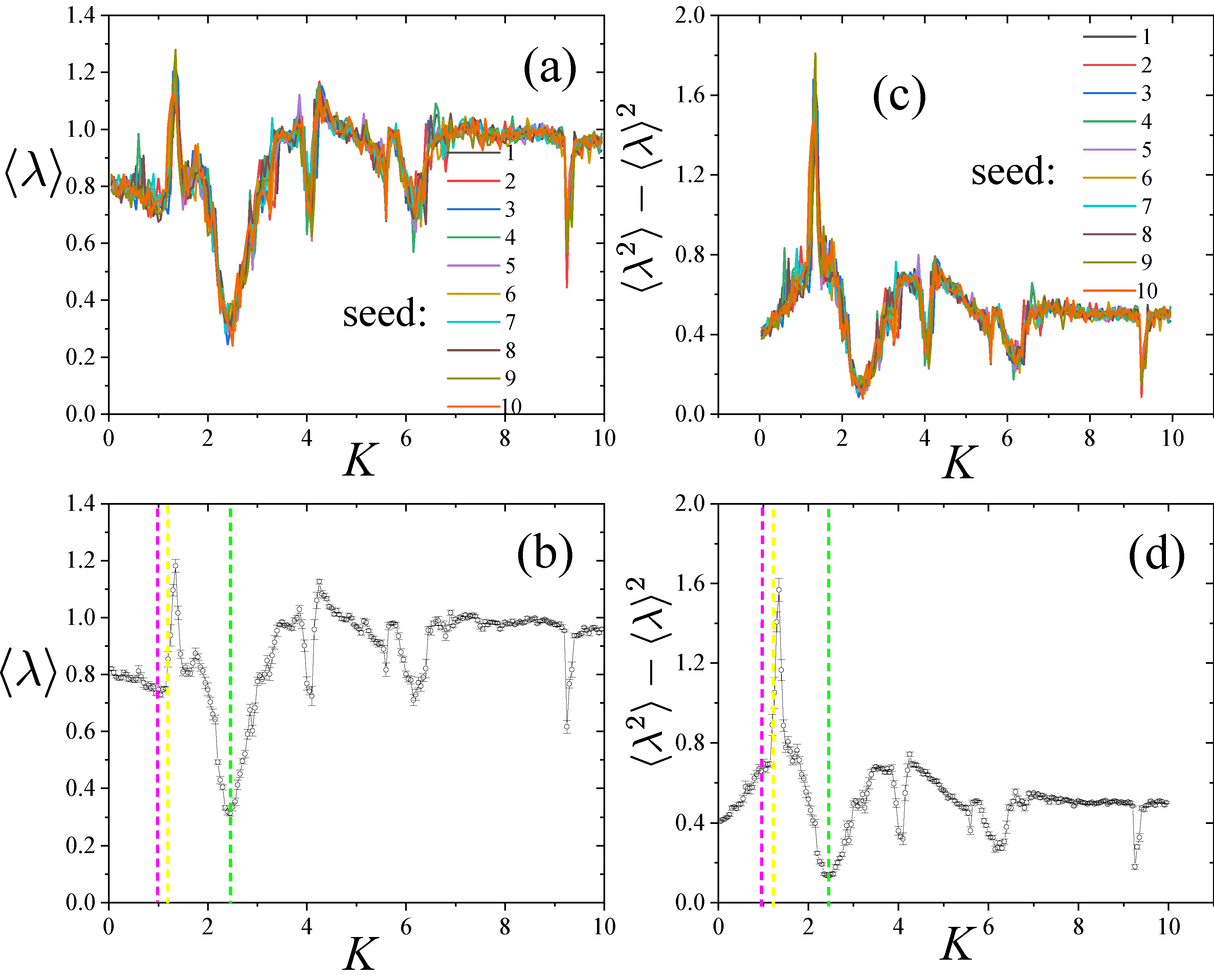}
\end{center}
\caption{Fluctuations in the eigenvalues of the matrix $G$ are investigated
by constructing it through the time evolutions of $p$.}
\label{Fig:fluctuations_p}
\end{figure}

These results are visually depicted in Figure \ref{Fig:fluctuations_p}. It
is quite intriguing to decipher the insights conveyed by this plot. In
Figure \ref{Fig:fluctuations_p} (a), the fluctuations in the average
eigenvalue are displayed as a function of $K$ for 10 different seeds.
Notably, the outcomes exhibit minimal variation across different seeds.
Figure \ref{Fig:fluctuations_p} (b) presents the same plot along with error
bars for added clarity.

\begin{figure}[tbp]
\begin{center}
\includegraphics[width=1.0\columnwidth]{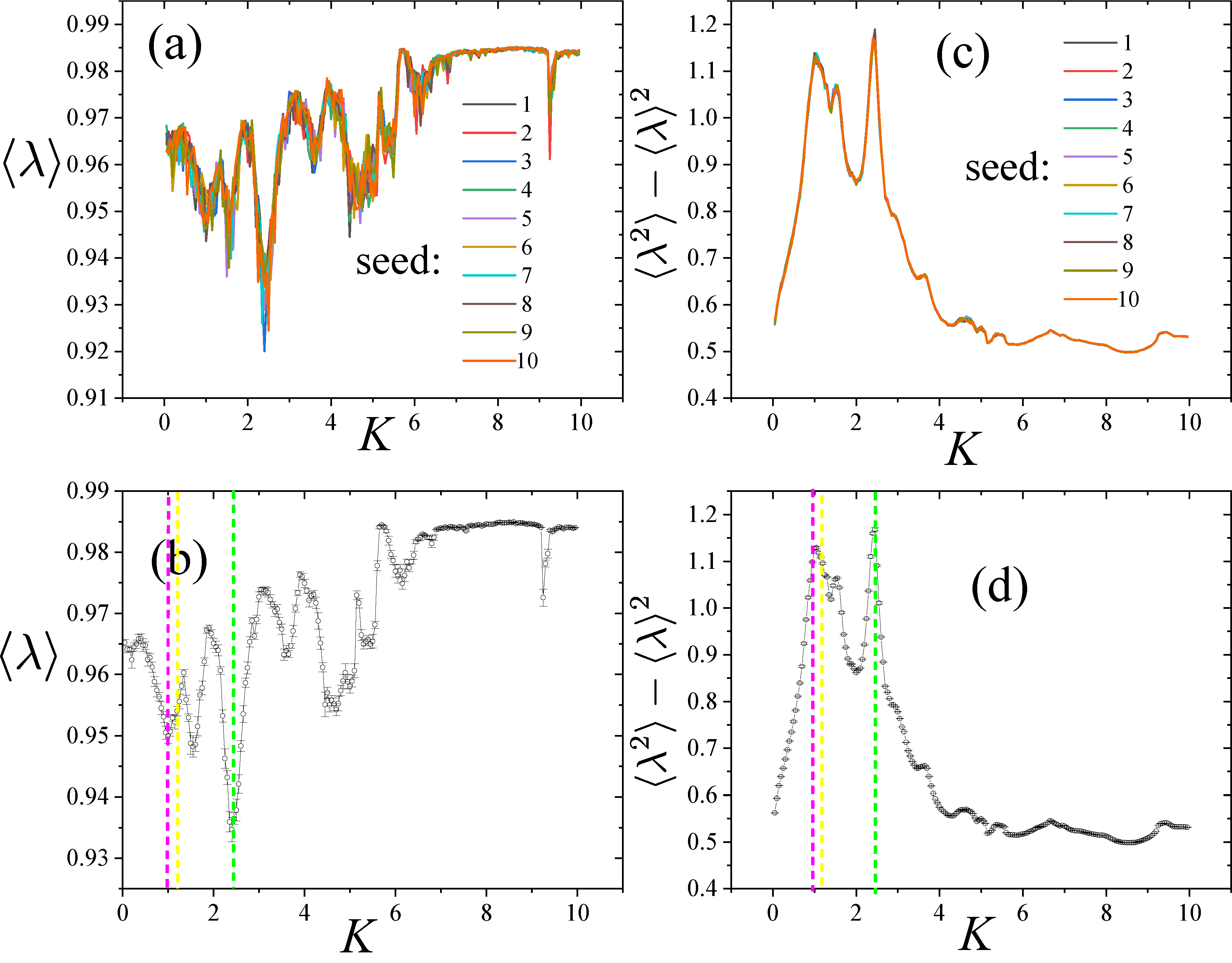}
\end{center}
\caption{The fluctuations in the eigenvalues of the matrix $G$, which is
constructed through iterations of $x$, are under examination.}
\label{Fig:fluctuations_x}
\end{figure}

We observe that the global minimum, denoted by the green dashed line,
precisely occurs at $K\approx 2.46$, not by coincidence. This numerical
value aligns with $\pi ^{2}/4$, which corresponds to the Chirikov
resonance-overlap criterion for the border of chaos. This same pattern
emerges in the dispersion of eigenvalues, as depicted in both Figure \ref%
{Fig:fluctuations_p} (c) and Figure \ref{Fig:fluctuations_p} (d). The former
illustrates the variance among different seeds, while the latter shows the
averaged under these seeds. Remarkably, $K\approx 2.46$ also serves as the
global minimum for the dispersion of eigenvalues.

The value $K\approx 2.46$ surpasses the baseline $K=$ $0.9716$ for several
reasons, including the influence of secondary-order resonances and the
finite width of the chaotic layer, as well as findings observed by \cite%
{Shepelyansky}. Interestingly, these authors note that such effects on the
Chirikov map appear to be less pronounced.

Furthermore, it is worth noting that $K=$ $0.9716$, indicated by the dashed
magenta line in the same plot, anticipates a significant upturn in both
behaviors, specifically in $\left\langle \lambda \right\rangle \times K$ and 
$\left\langle \lambda ^{2}\right\rangle -\left\langle \lambda \right\rangle
^{2}\times K$. This trend seems to exhibit universality. Lichtenberg and
Lieberman \cite{Lichtenberg} have proposed a refinement of $K$ as $K=\frac{%
\pi ^{2}}{4}$, which would suggest $K\approx 1.2$. This refinement is
represented by the yellow dashed line in Figure \ref{Fig:fluctuations_p}. It
is evident that this point consistently demonstrates an increase after the
critical $K=$ $0.9716$.

To provide a comprehensive view, we also examine the eigenvalues associated
with the time evolutions of positions. Interestingly, we continue to observe
the global minimum occurring at approximately $K\approx 2.46$, aligning with
the Chirikov resonance-overlap criterion for the chaos boundary. This
pattern appears to exhibit universality. However, a noteworthy departure
arises when considering the eigenvalue dispersion, as we now observe a
global maximum at the same $K$ value. This intriguing phenomenon is depicted
in Figure \ref{Fig:fluctuations_x}. To maintain consistency, we employ a
similar approach to illustrate the points $K=$ $0.9716$ and $K\approx 1.2$.

\section{Conclusions}

\label{Sec:Conclusions}

In our study, we utilize the technique of Wishart-like matrix spectra
fluctuations to probe the existence of chaos within Chirikov's standard map.
This methodology draws inspiration from its past success in characterizing
critical points within spin systems. Our results consistently affirm that
the resonance-overlap criterion for the chaos boundary, denoted as $K=\left( 
\frac{\pi }{4}\right) ^{2}\approx 2.46$, holds true whether we examine the
spectra obtained from the evolutions of moments or from the evolutions of
position coordinates.

However, intriguingly, when it comes to the dispersion, $\left\langle
\lambda ^{2}\right\rangle -\left\langle \lambda \right\rangle ^{2}$, this
same $K$ value remains a global minimum for the evolutions of moments but
transforms into a global maximum for the evolutions of positions.

Furthermore, the value $K=0.971635$, which marks the point at which the
golden KAM curve is disrupted, appears to foreshadow extreme behaviors in
eigenvalue fluctuations. This study holds promise in unraveling the
intricate relationship between chaos and quantum mechanics, but it merits
further exploration and extension to other models. Such analyses could prove
invaluable in enhancing our understanding of this complex interplay.


\begin{thebibliography}{99}
\bibitem{RMTFUTURA} We are currently engaged in research centered around the
utilization of random matrices as a tool for distinguishing between random
processes and chaotic behaviors. This research is currently in the
preparatory phase.
\end{thebibliography}
\end{document}